# The Stability of the Vacuum Polarization Surrounding a Charged Particle

F. J. Himpsel

Department of Physics, University of Wisconsin Madison,
1150 University Ave., Madison, WI 53706, USA, fhimpsel@wisc.edu

Abstract

The internal stability of the electron has been debated for a century at both the classical and the quantum level. Recently, a local force density balance was established for the 1$s$ electron in the H atom, based on the energy-momentum tensor of the classical Dirac field. This methodology is now extended to quantum fields by considering the force densities acting on the vacuum polarization induced by a point charge. Such a model is applicable to any charged particle at large distances, since the only vestige of its internal structure is the electric Coulomb field together with the vacuum polarization induced by it. While the polarization charge density is attracted to the point charge, it is kept from collapsing by repulsive forces due to confinement and degeneracy. It is shown analytically that the corresponding force densities are balanced for every filled shell of $m_j$ states at a given angular momentum $j$. The force densities are then summed over all single-electron states in the Dirac sea and renormalized by subtracting singular terms. In leading order of $\alpha$, the force densities remain balanced. This result establishes a local force balance for a prototypical manybody system.

## 1. Introduction

For more than a century the internal stability of the electron has preoccupied eminent physicists including Poincare, Lorentz, Fermi, Dirac, and Feynman [1],[2]. And for almost a century the fine structure constant $\alpha = e^2/\hbar c$ has posed a fascinating challenge to theorists [3]. Here we investigate the possibility of solving the two problems together by establishing a stability criterion for the vacuum polarization cloud surrounding a charged particle (including the electron). If a stability condition involves two opposing forces scaling with different powers of $\alpha$, the matching condition determines the value of the fine structure constant. Such a criterion was proposed by Casimir, but it did not work out quantitatively [4].

Recently the author established a criterion for the internal stability of the 1s electron in the H atom [5], using this prototypical system as testbed. It was found that the canonical formalism allows the definition of a local force density which is uniquely determined by the divergence of the stress tensor, the spatial part of the energy-momentum tensor. The continuity equation for the energy-momentum tensor ensures that the electrostatic attraction to the nucleus is automatically balanced by the repulsive force due to the confinement of the 1s electron to the Bohr radius. In that case the two forces scale the same power of $\alpha$, which implies that they are balanced for arbitrary values of $\alpha$.

This finding raised the question whether the automatic balance of the force densities is preserved in a manybody system, where renormalization alters the single-particle expressions. A prototypical manybody system for this purpose is the vacuum polarization surrounding a charged particle. The induced charge density is universal at large distances, i.e., independent of the internal structure of the charged particle, its mass, magnetic moment, and spin. In that limit it is dominated by electron-positron pairs, the lightest charged particles. The induced charge opposes the inducing charge and leads to electrostatic attraction – analogous to the H atom. Likewise, the balancing repulsive force is generated by the confinement of the vacuum polarization to the reduced Compton wavelength of the electron ($\lambdabar_C = \hbar/m_e c$). It is augmented by Pauli repulsion between the vacuum electrons (also known as degeneracy pressure). The latter is taken care of by summing over all singly-occupied states in the Dirac sea.

## 2. Force Densities Acting on the Electrons and Positrons in the Dirac Sea

The starting point is the symmetric energy-momentum tensor $T^{\mu\nu}$ for the Dirac field $\psi$ in the presence of an external electromagnetic field $A_\mu$ [6]:

(1) $\quad L_D = \{ \tfrac{1}{2} i [\bar{\psi}\gamma^\mu \cdot (\partial_\mu \psi) - (\partial_\mu \bar{\psi}) \cdot \gamma^\mu \psi] - m_e \bar{\psi}\psi \} - q(\bar{\psi}\gamma^\mu\psi) \cdot A_\mu \qquad q = -e \text{ for } e^-$

The opposing force densities $\mathbf{f}_\psi, \mathbf{f}_{A\psi}$ are obtained from the divergence of the the stress tensor $\mathbf{T}^{ij}$, the spatial part of $-T^{\mu\nu}$. With a static electric field $\mathbf{E}$ one obtains the following stress tensor and force densities for the Dirac field (see [5], including the Appendix):

(2) $\quad \mathbf{T}_\psi^{ij} = \tfrac{1}{4} i \{ [\bar{\psi}\gamma^i \cdot (\nabla_j \psi) - (\nabla_i \bar{\psi}) \cdot \gamma^j \psi] + [\bar{\psi}\gamma^j \cdot (\nabla_i \psi) - (\nabla_j \bar{\psi}) \cdot \gamma^i \psi] \}$

(3) $\quad \nabla \cdot \mathbf{T}_\psi = \mathbf{f}_\psi \qquad\qquad$ Confinement Force Density

(4) $\quad \rho \mathbf{E} = \mathbf{f}_{A\psi} \qquad\qquad$ Electrostatic Force Density

The charge density ρ of the Dirac field ψ is given by q·ψ*ψ, where q=±e is the charge of a positron (electron). The corresponding Dirac wave functions ψ are stationary and take the following form in the spherically-symmetric Coulomb potential $\Phi_C = e r^{-1}$ of a positive point charge:

(5) $\quad \psi_\kappa = \{ g_\kappa(r) \cdot \chi_\kappa^m(\theta,\varphi),\ i f_\kappa(r) \cdot \chi_{-\kappa}^m(\theta,\varphi) \} \cdot \exp(-iEt) \qquad \int \chi_\kappa^{m\dagger} \chi_\kappa^m d\Omega = 1$

$\quad \kappa = \pm 1, \pm 2, \pm 3, \ldots \qquad j = |\kappa| - \tfrac{1}{2} \qquad m_j = m \qquad l = |\kappa + \tfrac{1}{2}| - \tfrac{1}{2}$

In the following the label κ will be omitted from wave functions and related quantities. The radial functions f(r) and g(r) obey the following Dirac equations:

(6) $\quad f' = +\frac{\kappa-1}{r} \cdot f + [m_e - (E + \frac{\alpha}{r})] \cdot g$

$\quad g' = -\frac{\kappa+1}{r} \cdot g + [m_e + (E + \frac{\alpha}{r})] \cdot f$

The stress tensor $\mathbf{T}_\psi$ in (2) contains the tensor product of the vector $\{\gamma^r, \gamma^\theta, \gamma^\varphi\}$ with the gradient operator $\{\partial_r, r^{-1}\partial_\theta, (r\sin\theta)^{-1}\partial_\varphi\}$, applied to ψ. Its tensor divergence is the confinement force density $\mathbf{f}_\psi$. These quantities were calculated in [5] for the $s_{½}$ ground state in the H atom (see Equations (18),(19) therein). Here we need the expression for the complete range of angular momenta carried by the electrons and positrons in the Dirac sea. It turns out that only the $s_{½}$ and $p_{½}$ states (with κ=−1 and κ=+1) exhibit an isotropic force density. States with j>½ split into multiplets with different values of $m_j$. Each $|m_j|$ gives rise to an anisotropic force density. But the sum over a filled shell of $m_j$ produces again an isotropic result, multiplied by the degeneracy factor (2j+1)=2|κ|. This leads to the following stress tensor and the corresponding confinement force density of a filled $m_j$ shell ($m_j = -j, \ldots, j$):

(7) $\quad \Sigma_{mj} T_\psi^{rr} = -\frac{2|\kappa|}{4\pi} (fg' - f'g)$

$\quad \Sigma_{mj} T_\psi^{\theta\theta} = -\frac{2|\kappa|}{4\pi} \cdot \kappa \cdot fg/r$

$\quad \Sigma_{mj} T_\psi^{\varphi\varphi} = -\frac{2|\kappa|}{4\pi} \cdot \kappa \cdot fg/r$

(8) $\quad \Sigma_{mj} f_\psi^r = \frac{2|\kappa|}{4\pi} [\kappa \cdot 2fg/r^2 - 2(fg' - f'g)/r - (fg'' - f''g)]$

$\quad \Sigma_{mj} f_\psi^\theta = \Sigma_{mj} f_\psi^\varphi = 0$

The 4π in the denominator originates from the angular wave functions $\chi_\kappa^m$, which are normalized over the unit sphere.

The radial Coulomb force density of a filled shell is obtained in similar fashion. The sum over $m_j$ becomes isotropic and generates again the degeneracy factor 2|κ|. The combined charge density ρ is then simply multiplied with the radial electric field $E^r$:

(9) $\quad \Sigma_{mj} f_{A\psi}^r = \rho E^r = -\alpha \frac{2|\kappa|}{4\pi} (f^2 + g^2) \cdot r^{-2} \qquad \rho = -e \frac{2|\kappa|}{4\pi}(f^2+g^2) \qquad E^r = e r^{-2}$

One can now show analytically that the two force densities $\Sigma_{mj} f_{A\psi}^r$ and $\Sigma_{mj} f_\psi^r$ exactly compensate each other for every filled shell of $m_j$ at a given κ:

(10) $\quad \Sigma_{mj} f_\psi^r = \Sigma_{mj} f_{A\psi}^r$

This is accomplished by first taking the derivative of the radial Dirac equation in (6) in order to eliminate the 2nd derivatives g″,f″ from the expression for $\Sigma_{mj} f_\psi^r$ in (8). Then the radial Dirac equation itself is used to eliminate all remaining 1st derivatives from $\Sigma_{m_j} f^r_\psi$. After simplification one obtains the negative of $\Sigma_{m_j} f^r_{A\psi}$ in (9). This establishes an automatic force density balance for every complete shell of $m_j$ values. This balance holds for all discrete and continuum states, because the energy E does not appear in the expressions for the force densities.

Since the force density balance (10) exists for every filled shell, one is tempted to conclude that the force balance must also be satisfied after performing the summation over all vacuum states in the Dirac sea. That is not obvious, however, because the corresponding κ-sums over (8),(9) lead to linearly-diverging integrands for the integration over the momenta p of the continuum states. A well-defined summation and renormalization procedure needs to be established for obtaining a unique result. This is a characteristic of manybody systems as opposed to the single-electron system considered previously in [5].

## 3. Summation and Renormalization of the Force Densities

The summation over the infinite number of electrons and positrons in the Dirac sea involves a sum over negative and positive energies (= electron and positron states) and over all angular momenta $j=|\kappa|-½$ with $m_j=-j…+j$. This sum is combined with an integration over the radial momentum p for the continuum states or a sum over the principal quantum number n for discrete states. The latter will not be discussed, since they do not contribute to the vacuum polarization in lowest order of α [7]. The corresponding solutions of the Dirac equation in a Coulomb potential are given in Appendix A.

A direct summation over κ and integration over p was worked out in [7] for the vacuum polarization of $O(\alpha^3)$ and higher. Here we are interested in the leading term of $O(\alpha)$, the Uehling term [9]. In contrast to the higher orders it requires renormalization [8],[9]. The summation and the removal of infinities are described in Appendix B and tested against the analytic result obtained by standard methods [6]-[9]. After the κ-summation the momentum integration diverges quadratically and needs to be renormalized by a linear counter-term in the integrand. This term corresponds to the derivative of a δ-function in real space, which can be neglected when considering the force density at any finite distance. After the momentum integration a similar subtraction is required in real space to avoid a linear divergence of the vacuum polarization at large distances r. The result agrees with the analytic form of the vacuum polarization with an accuracy in the $10^{-5}$ range over two decades in r and three decades in the charge density. That is sufficient for establishing a force balance of $O(\alpha)$ for the leading term.

After testing the summation method for the vacuum polarization, one can extend it to the calculation of the electrostatic force density by simply multiplying each term in the vacuum polarization sum with the Coulomb field $E^r = e\,r^{-2}$. The corresponding sum for the confinement force density contains identical terms according to (10). Therefore the same summation scheme can be applied to both force densities, including the subtraction of singular terms. That establishes an overall force density balance in this manybody system.

## 4. Summary

This work establishes a local force balance for a prototypical manybody system, the vacuum polarization induced by a charged particle. The opposing force densities are electrostatic attraction and repulsion by confinement. It is instructive to compare this manybody system to a prototypical single-electron system, the H atom, where a similar force density balance was established recently (see Figure 1 in [5]). It involved the same canonical formalism for obtaining force densities from the divergence of the stress tensor. But there are interesting differences which can be seen by considering the dependence of the forces on the fine structure constant $\alpha$. The Bohr radius of the H atom shrinks with increasing $\alpha$, since it is proportional to $\alpha^{-1}$. As a consequence, the electron density increases, and with it the electrostatic attraction. The tighter confinement also increases the repulsive force, thereby keeping the forces balanced. The size of the vacuum polarization cloud is determined by the reduced Compton wavelength, which is independent of $\alpha$. Instead the number of induced electron-positron pairs increases when the electromagnetic coupling constant $\alpha$ becomes stronger. In a manybody system the number of particles is unlimited, while it remains fixed in a single-electron system, such as the H atom.

The vacuum polarization cloud around a point charge is of interest because this model system has universal character. It is applicable to the polarization charge surrounding any charged particle at large distances. Because of this universality, a stability criterion might open an avenue toward determining the electromagnetic coupling constant $\alpha$. That would simultaneously solve two of the most exciting physics problems from the last century, the stability of the electron and the calculation of $\alpha$. This concept works as long as the opposing forces scale with different powers of $\alpha$. The results reported here dampen these expectations somewhat, because the opposing forces are both proportional to $\alpha^2$ in leading order and thus generate an automatic force balance for all values of $\alpha$.

But the situation could change in higher orders. In fact, one could argue that any calculation of $\alpha$ should provide two solutions, i.e., the non-interacting case $\alpha=0$ and the observed value $\alpha \approx 1/137$. That requires at least a quadratic equation in $\alpha$ after canceling out the leading order. Consequently one would have to consider higher orders in $\alpha$. Such terms have been investigated for the vacuum polarization in muonic atoms and in atoms with high atomic number Z, where they cause significant effects [10]. Typically the potential energy $V = e \cdot \Phi$ is used to label the order of $\alpha$. The leading Uehling term becomes $\alpha(Z\alpha)$ and the next-to-leading term $\alpha^2(Z\alpha)$. After that come the terms $\alpha(Z\alpha)^3$ and $\alpha^2(Z\alpha)^2$. The corresponding electric fields $E^r$ and polarization charge densities $\rho$ are obtained via the gradient and the Laplacian of the potentials $\Phi$, respectively. Combining them pairwise into electrostatic force densities $\rho \cdot E^r$ produces a variety of electrostatic force densities. One can expect an equally rich set of confinement force densities which have yet to be determined. In higher orders of $\alpha$ the universality of the point charge model will be challenged. For example, the mass of a charged particle enters into the effective mass, which appears in the Dirac equation. There are also other forces to be considered, such as the exchange interaction of an electron with the virtual electrons in the vacuum of quantum electrodynamics [11].

**Appendix A: Notation, Solutions of the Dirac Equation**

The $(+---)$ signature is used for the metric. The unit system consists of $\hbar, c$, and the electron mass $m_e$, unless stated explicitly. The corresponding length unit is the reduced Compton wavelength $\lambdabar_C = \hbar/m_e c \approx 4 \cdot 10^{-13}$ m. Gaussian electromagnetic units yield the Coulomb potential $\Phi_C = e r^{-1}$ and the fine structure constant $\alpha = e^2/\hbar c \approx 1/137$. For other aspects of the notation, as well as for the Dirac equation in spherical coordinates see [5]. *Mathematica* 8 was used for analytic and numerical calculations.

**Discrete states with $|E|<m_e$:** The radial wave functions of the Dirac equation in the Coulomb potential consist of generalized Laguerre polynomials $L_n^\alpha(z)$ multiplied by a decaying exponential. The decay constant p is the imaginary part of the momentum:

(A1) $\quad E_{n,\kappa} = m_e \cdot [1 + \alpha^2/(n-\varepsilon)^2]^{-1/2} \quad n=1,2,3,... \quad m = (n-|\kappa|) \geq 0$

$$r \cdot g_{n,\kappa}(r) = p^{1/2} \cdot \tfrac{1}{2} \{ \alpha/[\kappa(\kappa-\gamma)]^{1/2} \cdot F^+_{n,\kappa}(r) + [(\kappa-\gamma)/\kappa]^{1/2} \cdot F^-_{n,\kappa}(r) \} \cdot \text{sign}(\kappa)$$

$$r \cdot f_{n,\kappa}(r) = p^{1/2} \cdot \tfrac{1}{2} \{ [(\kappa-\gamma)/\kappa]^{1/2} \cdot F^+_{n,\kappa}(r) + \alpha/[\kappa(\kappa-\gamma)]^{1/2} \cdot F^-_{n,\kappa}(r) \}$$

$\quad p = (m_e^2 - E_{n,\kappa}^2)^{1/2} \qquad \alpha > 0 \qquad \gamma = (\kappa^2 - \alpha^2)^{1/2} \quad \varepsilon = (|\kappa|-\gamma) \approx \alpha^2/2|\kappa|$

$\quad A^+_{n,\kappa} = [(m-1)!/\Gamma(m+2\gamma+1)]^{1/2} \cdot \{[(n-\varepsilon)+\gamma/\kappa \cdot (\alpha^2+(n-\varepsilon)^2)^{1/2}] \div [\gamma^2/\kappa^2 \cdot [\alpha^2+(n-\varepsilon)^2]]\}^{1/2}$

$\quad A^-_{n,\kappa} = \quad [m!/\Gamma(m+2\gamma)]^{1/2} \cdot \quad \{[(n-\varepsilon)-\gamma/\kappa \cdot (\alpha^2+(n-\varepsilon)^2)^{1/2}] \div [\gamma^2/\kappa^2 \cdot [\alpha^2+(n-\varepsilon)^2]]\}^{1/2}$

$\quad F^+_{n,\kappa}(r) = -A^+_{n,\kappa} \cdot (2pr)^{\gamma+1} \cdot L^{2\gamma+1}_{m-1}(2pr) \cdot \exp(-pr) \qquad F^+_{n,\kappa}(r) = 0 \text{ for } |\kappa|=n$

$\quad F^-_{n,\kappa}(r) = -A^-_{n,\kappa} \cdot (2pr)^{\gamma} \cdot L^{2\gamma-1}_{m}(2pr) \cdot \exp(-pr)$

$\quad \int [f^2_{n,\kappa}(r) + g^2_{n,\kappa}(r)] \cdot r^2 \, dr = 1 \qquad \int \psi^*_{n',\kappa',mj'}(\mathbf{r}) \cdot \psi_{n,\kappa,mj}(\mathbf{r}) \, d^3\mathbf{r} = \delta_{n,n'} \cdot \delta_{\kappa,\kappa'} \cdot \delta_{mj,mj'}$

**Continuum states with $|E|>m_e$:** The radial Coulomb solutions contain the real and imaginary part of the regular Whittaker function $M_{\kappa,\mu}$, taken for imaginary momenta $i2p$:

(A2) $\quad r \cdot g_{p,\kappa}(r) = +A \cdot [(E+m_e)/2E]^{1/2} \cdot (2pr)^\gamma \cdot \text{Re}[F(\alpha,\kappa,p,r)]$

$\qquad r \cdot f_{p,\kappa}(r) = -A \cdot [(E-m_e)/2E]^{1/2} \cdot (2pr)^\gamma \cdot \text{Im}[F(\alpha,\kappa,p,r)] \cdot \text{sign}(E)$

$\quad E = \pm(m_e^2+p^2)^{1/2} \qquad \gamma = (\kappa^2-\alpha^2)^{1/2} \qquad \delta = \alpha \cdot E/p$

$\quad A = (2/\pi)^{1/2} \cdot |\Gamma(\gamma+i\delta)|/\Gamma(1+2\gamma) \cdot \exp(\tfrac{\pi}{2}\delta)$

$\quad F(\alpha,\kappa,p,r) = [(-\kappa+i\alpha/p) \cdot (\gamma+i\delta)]^{1/2} \cdot (i2pr)^{-(\gamma+1/2)} \cdot M_{-(1/2+i\delta),\gamma}(i2pr)$

$\quad r^2 \cdot [f^2_{p,\kappa}(r) + g^2_{p,\kappa}(r)] \to \pi^{-1} \quad \text{for} \quad pr \to \infty$

$\qquad \int \psi^*_{p',\kappa',mj'}(\mathbf{r}) \cdot \psi_{p,\kappa,mj}(\mathbf{r}) \, d^3\mathbf{r} = \delta(p-p') \cdot \delta_{\kappa,\kappa'} \cdot \delta_{mj,mj'}$

$\qquad \Sigma_{\kappa,mj} \int \psi^*_{p,\kappa,mj}(\mathbf{r}) \cdot \psi_{p,\kappa,mj}(\mathbf{r}') \, dp = \delta^3(\mathbf{r}-\mathbf{r}')$

In the limit $\alpha \to 0$ one obtains the free-electron solutions containing Bessel functions:

(A3) $\quad r \cdot g_{p,\kappa}(r) = -[(E+1)/2E]^{1/2} \cdot (pr)^{1/2} \cdot J_{|+\kappa+1/2|}(pr) \cdot \text{sign}(\kappa)$

$\qquad r \cdot f_{p,\kappa}(r) = -[(E-1)/2E]^{1/2} \cdot (pr)^{1/2} \cdot J_{|-\kappa+1/2|}(pr) \cdot \text{sign}(E)$

## Appendix B: Vacuum Polarization by Direct Summation

The vacuum polarization potential of a point charge $+e$ is known analytically in lowest order of $\alpha$ [6]-[9]. The corresponding charge density can be derived via the Laplacian and expressed in terms of modified Bessel and Struve functions $K_n$ and $\mathbf{L}_m$:

(B1) $\quad \rho_{VP} = -e\,\frac{2\alpha}{3\pi^2}\,m_e^3 \cdot \{\, Ki_1(z) + z^{-1}\cdot K_0(z) - (1 - 2z^{-2})\cdot K_1(z)]\,\} \qquad z = 2m_e r$

$\qquad\qquad Ki_1(z) = \int_z^\infty K_0(y)\,dy = \pi/2 \cdot \{\,1 - z\cdot[K_0(z)\cdot \mathbf{L}_{-1}(z) - K_1(z)\cdot \mathbf{L}_0(z)\,]\,\}$

$\qquad \rho_{VP} \to -e\,\sqrt{2}\,\frac{\alpha}{8\pi^{3/2}}\cdot r^{-3}\cdot z^{1/2}\,e^{-z} \quad$ for $r \to \infty$

$\qquad \rho_{VP} \to -e\,\frac{\alpha}{6\pi^2}\cdot r^{-3} \qquad\qquad\qquad$ for $r \to 0$

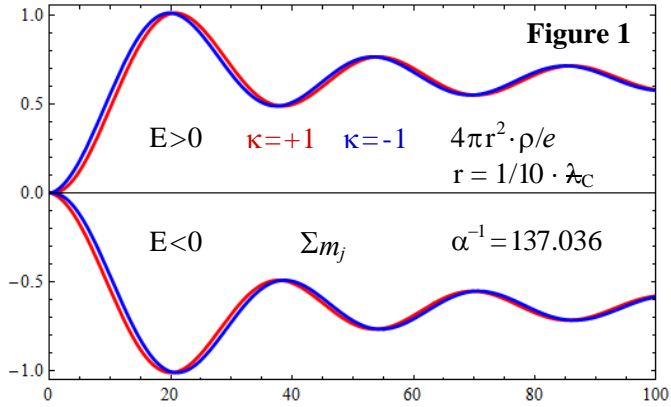

**Figure 1**

$E>0 \quad \kappa=+1 \quad \kappa=-1 \quad 4\pi r^2 \cdot \rho/e$
$\qquad\qquad\qquad\qquad\qquad r = 1/10 \cdot \lambdabar_C$
$E<0 \quad \Sigma m_j \quad\quad \alpha^{-1} = 137.036$

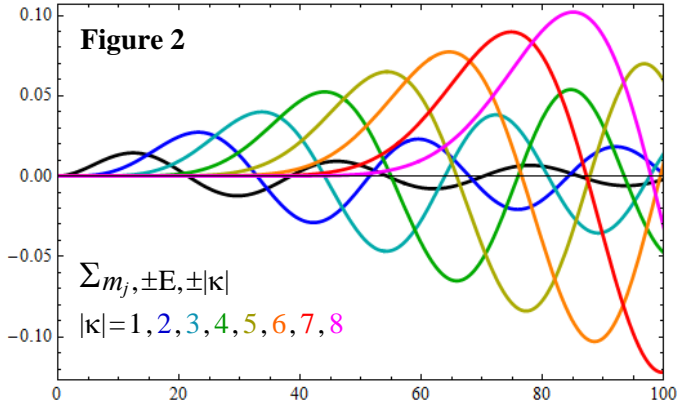

**Figure 2**

$\Sigma m_j, \pm E, \pm|\kappa|$
$|\kappa| = 1, 2, 3, 4, 5, 6, 7, 8$

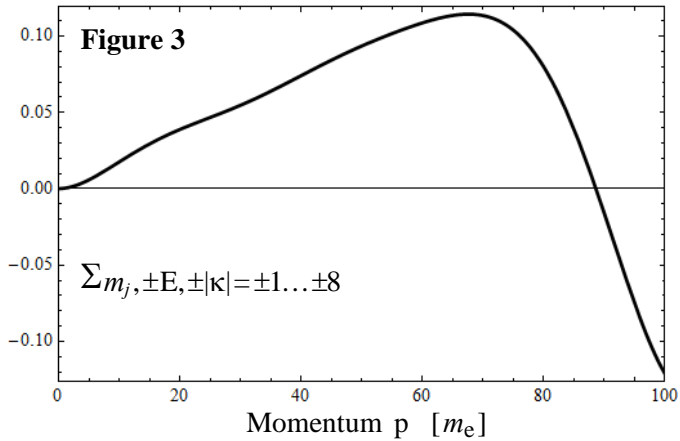

**Figure 3**

$\Sigma m_j, \pm E, \pm|\kappa| = \pm 1 \ldots \pm 8$

Momentum p [$m_e$]

This O($\alpha$) result for $\rho_{VP}$ is used to test the direct summation method, with the caveat that the sum also contains higher order terms [7].

The following Figures 1-9 illustrate the sequence of summations for the charge density $\rho$ (multiplied by $4\pi r^2/e$). It starts with the charge-symmetric average over electrons ($\rho_-, E<0$) and positrons ($\rho_-, E>0$):

(B2) $\quad \rho = \tfrac{1}{2}(\rho_- - \rho_+)$

$\rho_\pm$ are defined in (9), with $f, g$ given in (5),(A2). The sum over all $m_j$ states at a given $\kappa$ has already been taken in (9). The charge densities for $\pm E, \pm\kappa$ are combined right away, since they nearly compensate each other (Fig. 1 and Fig. 2, black line).

**Figure 1**: Electron states (E<0) and positron states (E>0) nearly cancel each other, leaving a sum of O($\alpha$).

**Figure 2**: Charge densities summed over $\pm E, \pm \kappa$. They shift up in p with increasing $|\kappa|$ and thereby lead to a converging $\kappa$-sum at fixed p. The black curve is the sum over Fig. 1.

**Figure 3**: This finite sum over eight $|\kappa|$ has converged in the left half, but not yet on the right. The maximum value of $|\kappa|$ required for convergence increases with $p\cdot r$.

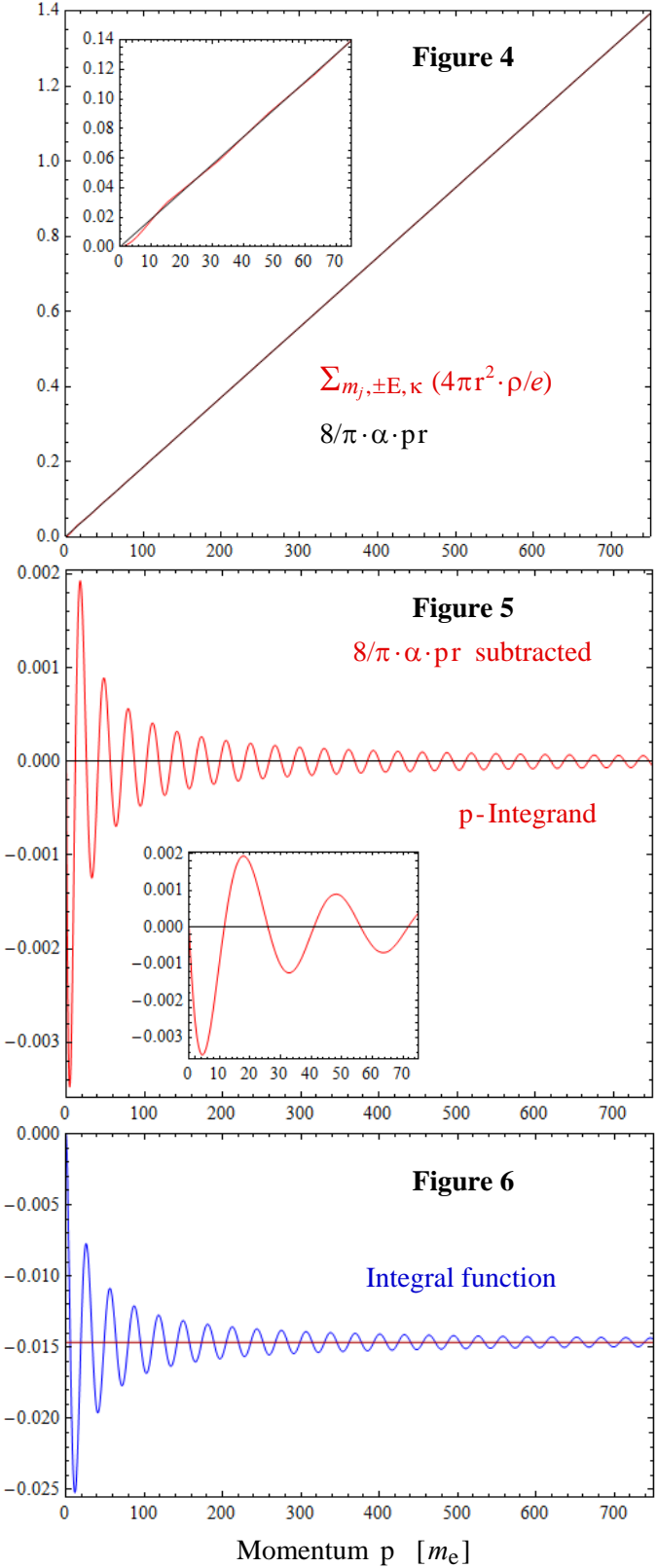

**Figure 4**: A converged κ-sum over a wider range of p (red line). It is almost identical to the straight line $8\alpha/\pi \cdot r$ (black). The inset reveals small oscillations about the black line near $p=0$.

**Figure 5**: The charge density oscillations in Figure 4 show up clearly after subtracting the straight line $8\alpha/\pi \cdot pr$. Its Fourier transform from p to r has the form $r \cdot \delta'(r)$, which does not contribute for $r>0$.

**Figure 6**: The integral function of the charge density in Figure 5. It converges toward a finite, well-defined value at large p (red line). The convergence of the oscillating integral function is accelerated by a simple averaging method which dampens the oscillations by several orders of magnitude (discussed with Figure 10). The remaining oscillations are reduced further by taking values at a series of inflection points and averaging over pairs of inflection points with opposite slopes. For $r>\frac{1}{2}$ this series converges like $-\alpha/2\pi \cdot r/p^2$ toward the limit of the integral function.

Such sums over $m_j, \pm|\kappa|, \pm E$, and p are calculated for r-points ranging from $4\lambda_C$ down to $\lambda_C/100$ where virtual muons and pions begin to contribute. $\rho_{VP}$ decays exponentially above $\lambda_C$, creating quickly numerical problems. At small r it increases with a power law as given in (B1). The result of the p-integration is the induced charge density as a function of r, which is shown in Figures 7-9.

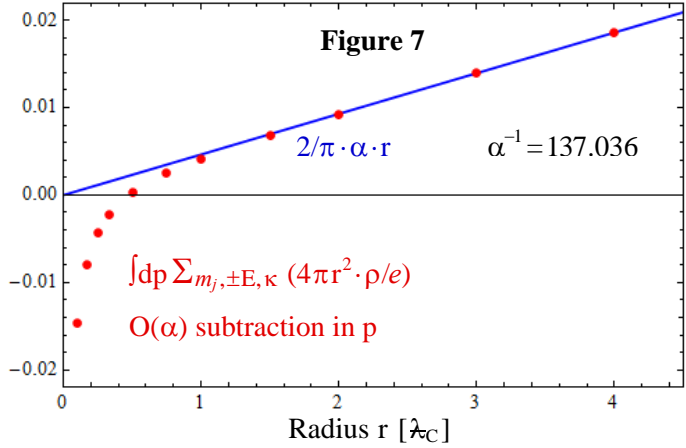

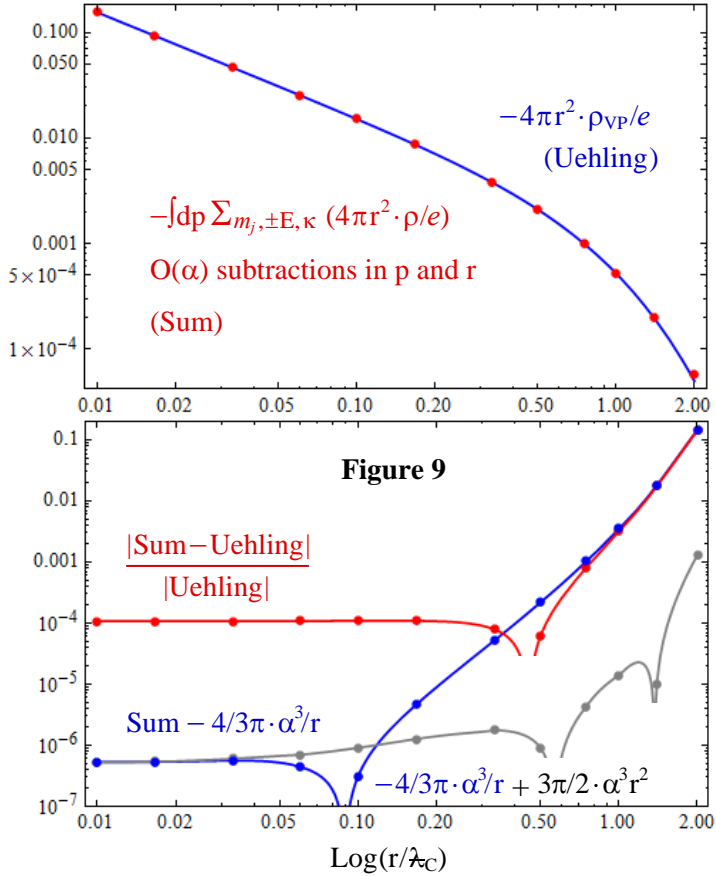

**Figure 7**: The induced charge density, summed over $m_j, \pm E, \kappa$, integrated over p, and plotted versus r (dots). The linear increase at large r requires a second subtraction to ensure that the charge density converges to zero for large r. The subtraction term $2\alpha/\pi \cdot r$ has a singular Fourier transform which is concentrated at $p=0$.

**Figure 8**: Comparison of the analytic $O(\alpha)$ result (B1) for the vacuum polarization (line) with the result obtained here by summation over the Dirac sea and subtraction of singular terms (dots). There is a match over several decades in r and $\rho_{VP}$.

**Figure 9**: Normalized difference between the sum over the Dirac sea (Sum) and the analytic $O(\alpha)$ result (Uehling). They agree at the $10^{-6}$ level for $r<½$ after subtracting two terms of $O(\alpha^3)$ from the sum. The contact term $4/3\pi \cdot \alpha^3/r$ is required when summing over $\kappa$ before integrating over p [7]. The term $-3\pi/2 \cdot \alpha^3 r^2$ is singular in p, analogous to the $O(\alpha)$ term in Fig. 7. Inverted cusps occur at sign changes.

These results were obtained by first carrying out the $\kappa$-summation on a grid of $>10^3$ p-points with spacing proportional to $1/r$ (finer at small p where the dominant part of the integral is located). The resulting p-integrand was interpolated by 7$^{th}$ order polynomials, which allowed analytic integration over p, as well as analytic differentiation to find the inflections points of the integral function. The upper limit of the p-integration was typically $p_{max}=160/r$, which required summations up to $|\kappa|\approx 200$ when approaching $p_{max}$. After making subtractions of $O(\alpha^3)$ the resulting precision becomes much better than needed for $O(\alpha)$. Subtracting the additional $O(\alpha^5)$ term $4\pi^2 \cdot \alpha^5/r$ reveals the $O(\alpha^3)$ vacuum polarization (not shown). The $O(\alpha^2)$ term is not included in this summation [7].

One of the obstacles for a straight summation of individual charge densities is the extraction of precise integrals from oscillating integrands that are weakly damped. For this purpose an efficient method for damping the oscillations of a nearly-periodic function F(z) was used. Let $p$ be the period of F. The oscillations oppose each other when phase-shifting F(z) by a half-period $\delta z = \frac{1}{2}p$ and averaging between F(z) and $F(z+\frac{1}{2}p)$. Performing the analogous average with $\delta z = -\frac{1}{2}p$ and averaging over the two averages dampens the oscillations by several orders of magnitude:

(B3)  $\overline{F}(z) = \frac{1}{4}F(z-\frac{1}{2}p) + \frac{1}{2}F(z) + \frac{1}{4}F(z+\frac{1}{2}p)$

For the integrand of the vacuum polarization charge density the period becomes $p=\pi/r$. The effect of this averaging procedure is illustrated in Figure 10, using the tail of the integral function from Figure 6 as example.

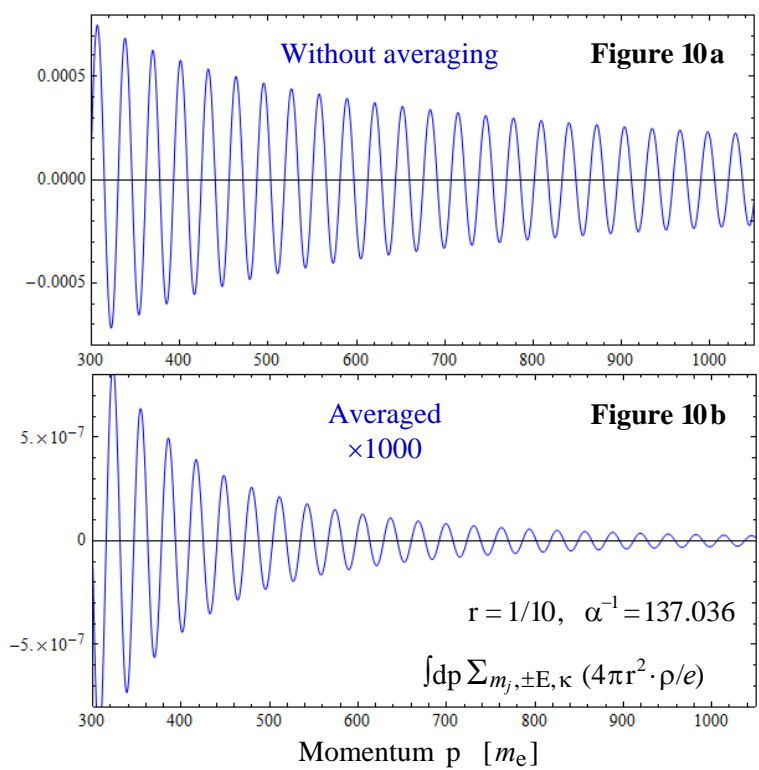

**Figure 10**: Oscillations of the integral function from Figure 6 around its asymptotic value (which has been subtracted). The averaging method defined in (B3) reduces the oscillation amplitude from a) to b) by more than 3 orders of magnitude. The effect is particularly strong at the upper cutoff of the p-integral, where the phase shift of the Coulomb wave functions becomes less dependent on p. See also the caption of Fig. 6 for further damping and extrapolation toward $p \to \infty$.